\documentclass[aps,showpacs,showkeys,preprintnumbers,amsmath,amssymb]
{revtex4}
\usepackage{graphicx}
\begin{document}

\title{Vacuum fluctuation effects on hyperonic neutron star matter}
\author{ M.-H. Weng$^{1,}$\footnote[1]{ \emph{E-mail address:} mhweng@mail.bnu.edu.cn},
 X.-H. Guo$^{1,}$\footnote[2]{Corresponding author, \emph{E-mail address:}
 xhguo@bnu.edu.cn},
 and B. Liu$^{2,}$\footnote[3]{Corresponding author, \emph{E-mail address:} liub@mail.ihep.ac.cn}}
\affiliation{$^{1}$ College of Nuclear Science and Technology, Beijing Normal University, Beijing 100875, People's Republic of China\\
$^{2}$ Institute of High Energy Physics, Chinese Academy of
Sciences, Beijing 100049, People's Republic of China}

%%%%%%%%%%%%%%%%%%%%%%%%%%%%%%%%%%%%%%%%%%%%%%%%%%%%%%

\begin{abstract}

The vacuum fluctuation (VF) effects on the properties of the
hyperonic neutron star matter are investigated in the framework of
the relativistic mean field (RMF) theory. The VF corrections result
in the density dependence of in-medium baryon and meson masses. We
compare our results obtained by adopting three kinds of
meson-hyperon couplings. The introduction of both hyperons and VF
corrections soften the equation of state (EoS) for the hyperonic
neutron star matter and hence reduce hyperonic neutron star masses.
The presence of the $\delta$ field enlarges the masses and radii of
hyperonic neutron stars. Taking into account the uncertainty of
meson-hyperon couplings, the obtained maximum masses of hyperonic
neutron stars are in the range of $1.33M_{\odot}\sim1.55M_{\odot}$.

\end{abstract}



\maketitle

\section*{I. Introduction}

The nonlinear Walecka model (NLWM), based on the relativistic mean
field (RMF) theory, has been successfully used in the study of
nuclear matter and neutron stars
\cite{wal,bri,sug,mue95,mue96,liu02,liu05}. Neutron stars are
composed of highly compressed matter. Nuclear matter at high
densities exhibits a new degree of freedom: strangeness. Hyperons,
kaon condensation, and quarks may appear in neutron stars, and these
complicated compositions of neutron stars have attracted much
attention. The important property of a neutron star is characterized
by its mass and radius, which can be obtained from the appropriate
equation of state (EoS) at high densities. The RMF model was first
used to investigate the properties of hyperonic neutron stars
($npe\mu H$) ($H$ denotes hyperons throughout this paper) in 1980s
\cite{gle82,gle85}. Recently, in some studies based on the RMF model
it has been indicated that the hyperons play an important role in
the neutron star matter \cite{she,esp,men}. In recent years it has
been stressed that the inclusion of the $\delta$ field is important
in the study of the asymmetric nuclear matter
\cite{liu02,liu05,men,gai,bar05}. The inclusion of the $\delta$
field leads to the structure of relativistic interactions, where a
balance between an attractive (scalar) and a repulsive (vector)
potential exists. The $\delta$ field plays a role in the isospin
channel and mainly affects the behavior of the system in the high
density regions and so is of great interest in nuclear astrophysics.
The influence of the $\delta$ field on the properties of hyperonic
neutron stars has been investigated based on the RMF model
\cite{sch,men,mi}.

In Ref. \cite{bha}, the vacuum fluctuation (VF) corrections were
taken into account to study the properties of nuclear matter. In our
recent paper \cite{guo}, we developed the VF-RMF model by including
the isovector mesons ($\rho$ and $\delta$) to investigate the
properties of the asymmetric nuclear matter and neutron stars. The
VF effects lead to the dependence of in-medium hadron masses on the
total baryon density. In this work, we will extend the VF-RMF model
to hyperon-rich matter in neutron stars by including the hyperons
and leptons in the relativistic Lagrangian density. The VF effects
will be introduced by considering loop corrections in the
self-energies of in-medium baryons and mesons as in Ref. \cite{guo}.
The VF effects on the properties of the hyperonic neutron star
matter will be studied.

This article is organized as follows. In Sec. II, we derive the
in-medium masses of baryons and mesons and the EoS for the hyperonic
neutron star matter in the VF-RMF model. Sec. III is devoted to our
results and discussions. In Sec. IV, a brief summary is presented.

\section*{II. The baryon octet VF-RMF model}
{\hskip  0.7cm}

The relativistic Lagrangian density with the baryon octet and free
leptons used in this work reads
\begin{eqnarray}\label{eq:1}
{\cal L } &=& \sum_{B}\bar{\psi}_B [i\gamma_{\mu}\partial^{\mu}
-(M_B - g_{\sigma B}\phi - g_{\delta B}\vec{t}_{B}\cdot\vec{\delta})
-g_{\omega B}\gamma_\mu\omega^{\mu}-g_{\rho
B}\gamma^{\mu}\vec{t}_{B}\cdot \vec{b}_{\mu}]\psi_B \nonumber \\&&
+\frac{1}{2}(\partial_{\mu}\phi\partial^{\mu}\phi-m_{\sigma}^2\phi^2)
-U(\phi)+\frac{1}{2}m^2_{\omega}\omega_{\mu} \omega^{\mu}
+\frac{1}{2}m^2_{\rho}\vec{b}_{\mu}\cdot\vec{b}^{\mu} \nonumber\\&&
+\frac{1}{2}(\partial_{\mu}\vec{\delta}\cdot\partial^{\mu}\vec{\delta}
-m_{\delta}^2\vec{\delta^2}) -\frac{1}{4}F_{\mu\nu}F^{\mu\nu}
-\frac{1}{4}\vec{G}_{\mu\nu}\vec{G}^{\mu\nu}\nonumber\\&&
+\sum_{l}\bar{\psi}_l(i\gamma_{\mu}\partial^{\mu}-m_l)\psi_l+
\delta{\cal L},
\end{eqnarray}
where the sum on $B$ is over all the states of the lowest baryon
octet (with mass $M_{B}$)
 ($B$=$n$, $p$, $\Lambda$, $\Sigma^{-}$, $\Sigma^{0}$, $\Sigma^{+}$, $\Xi^{-}$, $\Xi^{0}$)
 and the sum on $l$ is over the free leptons (with mass $m_{l}$) ($l$=$e^{-}$, $\mu^{-}$);
  $\phi$, $\omega_{\mu}$, $\vec{b}_{\mu}$, and
$\vec{\delta}$ (with masses $m_{\sigma}$, $m_{\omega}$, $m_{\rho}$,
$m_{\delta}$, respectively) represent $\sigma$, $\omega$, $\rho$,
and $\delta$ meson fields, respectively; $\vec{t}_{B}$ represents
the isospin generator matrix for the baryon $B$;
$U(\phi)=\frac{1}{3}a\phi^{3}+\frac{1}{4}b\phi^{4}$ is
 the nonlinear potential of the $\sigma$ meson,
$F_{\mu\nu}\equiv\partial_{\mu}\omega_{\nu}-\partial_{\nu}\omega_{\mu}$
and
$\vec{G}_{\mu\nu}\equiv\partial_{\mu}\vec{b}_{\nu}-\partial_{\nu}\vec{b}_{\mu}$;
the counterterm for the Lagrangian density, $\delta{\cal L}$, has
the same form as that in Ref. \cite{guo}.

The field equation for the baryon $B$ in the RMF approximation is
given by
\begin{eqnarray}\label{eq:2}
&& [i\gamma_{\mu}\partial^{\mu}-(M_B - g_{\sigma B}\phi -g_{\delta
B}{t_{3B}}\delta_3) -g_{\omega B}\gamma^{0}{\omega_0} -g_{\rho
B}\gamma^{0}{t_{3B}}{b_0}]\psi_B=0,
\end{eqnarray}
with
\begin{eqnarray}\label{eq:3}
&& \phi=\frac{1}{m_{\sigma}^2}( \sum_{B} g_{\sigma B}\rho_{sB}-a\phi^2-b\phi^3), \nonumber \\
&& \omega_{0}=\frac{1}{m_{\omega}^2}\sum_{B} g_{\omega B}\rho_B,\nonumber \\
&& b_{0}=\frac{1}{m^2_{\rho}} \sum_{B} g_{\rho B}t_{3B}\rho_B,\nonumber \\
&& \delta_3=\frac{1}{m^2_{\delta}}\sum_{B} g_{\delta B}
t_{3B}\rho_{B}^{s},
\end{eqnarray}
where $t_{3B}$ is the third direction projection of the
$\vec{t}_{B}$ for baryon $B$. $\rho_B$ and $\rho_{B}^{s}$ are
the number and scalar densities of the baryon  $B$, which are
given in the following respectively,
\begin{eqnarray}\label{eq:4}
\rho_{B}&=&\frac{k_{F_{B}}^{3}}{3\pi^2},
\end{eqnarray}
and
\begin{eqnarray}\label{eq:5}
\rho_{B}^{s}&=& - i\int \frac{{\rm d} ^4k}{(2\pi)^4}{\rm
Tr}G^{B}(k),
\end{eqnarray}
where $k_{F_{B}}$ is the Fermi momentum of the baryon $B$ and
$G^{B}(k)$ is the propagator of the baryon $B$ in the VF-RMF
model:
\begin{eqnarray}\label{eq:6}
G^{B}(k)&=&(\gamma_{\mu}k^{\mu}+M_{B}^{\star})\Bigl[\frac{1}{k^2-M_{B}^{\star2}+i\eta}
+\frac{i\pi}{E_{F_{B}}^{\star}}\delta(k^0-E_{F_{B}})\theta(k_{F_{B}}-|\vec{k}|)\Bigr]\nonumber\\
&\equiv&G^{B}_{F}(k)+G^{B}_{D}(k),
\end{eqnarray}
where $M_{B}^{\star}$ is the effective mass of the baryon  $B$,
$E_{F_{B}}^{(\star)}=\sqrt{k_{F_{B}}^2+M_{B}^{(\star)2}}$ and $\eta$
is infinitesimal.

\begin{figure}[hbtp]\label{fig01}
\begin{center}
\includegraphics[scale=0.8]{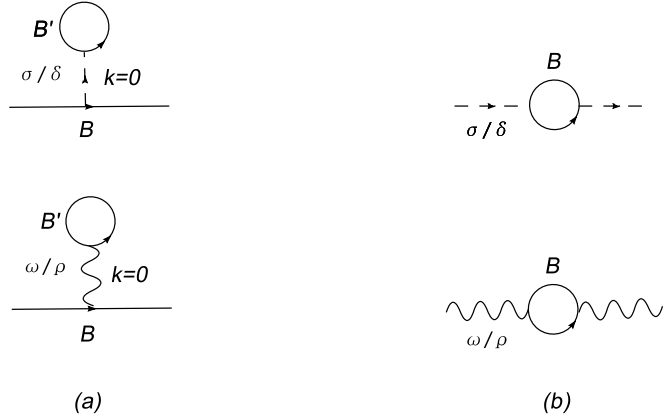}
\caption{Loop-diagram corrections to the self-energy of baryon octet
states (a) and mesons (b) in medium, where $B^{(\prime)}$ denotes
baryons and
 $k$ is the four momentum of the meson.}
\end{center}
\end{figure}

In the present work, we only consider the dominant VF contributions
from the tadpole diagrams to the self-energies of baryon octet
states. Thus, when the VF corrections are introduced through Fig.
1(a), the effective mass of the baryon  $B$ can be written as:
\begin{eqnarray}\label{eq:7}
  M_{B}^{\star}&=&M_{B}+ig_{\sigma B}\sum\limits_{B^{\prime}}\frac{g_{\sigma B^{\prime}}}{m_{\sigma}^{\star 2}}\int \frac{{\rm d} ^4k}{(2\pi )^4}{\rm
 Tr}G^{B^{\prime}}(k)\nonumber\\
 &&+ig_{\delta B}t_{3B}\sum\limits_{B^{\prime}}\frac{g_{\delta B^{\prime}}}{m_{\delta}^{\star 2}}t_{3B^{\prime}}\int \frac{{\rm d} ^4k}{(2\pi )^4}{\rm
 Tr}G^{B^{\prime}}(k)\nonumber\\
 &&+\frac{g_{\sigma B}}{m_{\sigma}^{\star2}}(a\phi^{2}+b\phi^{3}),
\end{eqnarray}
where $m_{j}^{\star}$ ($j=\sigma, \omega, \rho, \delta$ throughout
this paper) are the off-shell in-medium meson masses.

The introduction of the density dependence of the in-medium meson
masses is the critical effect of VF corrections. Because the meson
propagators in the baryon self-energies carry zero four-momenta, we
must use the off-shell $(q^{\mu}=0)$ meson masses in the tadpole
loop calculations for self consistency. We calculate the in-medium
meson masses in the random-phase approximation (RPA) \cite{bha,kur},
see Fig. 1(b).
The obtained off-shell effective mass of the $\sigma$ meson is given by
\begin{equation}\label{eq:8}
 m_{\sigma}^{\star2}=m_{\sigma}^{2}+\Pi_{\sigma}(q^{\mu}=0),
\end{equation}
where
\begin{equation}\label{eq:9}
\Pi_{\sigma}(q^{\mu}=0)=-i\sum\limits_{B}g_{\sigma
B}^{2}\int\frac{{\rm d^4}k}{(2\pi)^4}{\rm Tr} [G^{B}(k+q)G^{B}(k)].
\end{equation}

The off-shell effective mass of the $\delta$ meson is obtained as
follows:
\begin{equation}\label{eq:10}
 m_{\delta}^{\star2}=m_{\delta}^{2}+\Pi_{\delta}(q^{\mu}=0),
\end{equation}
where
\begin{equation}\label{eq:11}
\Pi_{\delta}(q^{\mu}=0)=-i\sum\limits_{B}g_{\delta
B}^{2}t_{3B}^{2}\int\frac{{\rm d^4}k}{(2\pi)^4}{\rm Tr}
[G^{B}(k+q)G^{B}(k)].
\end{equation}

The off-shell effective mass of the $\omega$ meson is given by
\begin{equation}\label{eq:12}
 m_{\omega}^{\star2}=m_{\omega}^{2}+\Pi_{\omega T}(q^{\mu}=0),
\end{equation}
where $\Pi_{\omega T}$ is the transverse part of the following
polarization tensor:
\begin{equation}\label{eq:13}
\Pi_{\omega}^{\mu\nu}(q^{\mu}=0)=-i\sum\limits_{B}g_{\omega
B}^{2}\int\frac{{\rm
 d^4}k}{(2\pi)^4}{\rm Tr}
 [\gamma^{\mu}G^{B}(k+q)\gamma^{\nu}G^{B}(k)].
\end{equation}

The off-shell effective mass of the $\rho$ meson is given by
\begin{equation}\label{eq:14}
 m_{\rho}^{\star2}=m_{\rho}^{2}+\Pi_{\rho T}(q^{\mu}=0),
\end{equation}
where $\Pi_{\rho T}$ is the transverse part of the following
polarization tensor:
\begin{equation}\label{eq:15}
\Pi_{\rho}^{\mu\nu}(q^{\mu}=0)=-i\sum\limits_{B}g_{\rho
B}^{2}t_{3B}^{2}\int\frac{{\rm
 d^4}k}{(2\pi)^4}{\rm Tr}
[\gamma^{\mu}G^{B}(k+q)\gamma^{\nu}G^{B}(k)].
\end{equation}

Obviously,  the modification of the in-medium hadron masses will
affect the properties of the hyperonic neutron star matter. The
meson masses appearing in the Lagrangian density should be replaced
by the off-shell in-medium meson masses in our calculations.
Therefore, the energy-momentum tensor in the VF-RMF model can be
expressed as
\begin{equation}\label{eq:16}
T_{\mu\nu}=\sum_{B}i\bar{\psi}_B \gamma_{\mu}\partial_{\nu}\psi_B
+\sum_{l}i \bar{\psi}_l \gamma_{\mu}\partial_{\nu}\psi_l
+g_{\mu\nu}\Bigl[\frac{1}{2} m_{\sigma}^{\star2}\phi^2 +U(\phi)
-\frac{1}{2}m^{\star2}_{\omega}\omega_{\lambda} \omega^{\lambda}
-\frac{1}{2}m^{\star2}_{\rho}\vec{b_{\lambda}}\vec{b^{\lambda}}
+\frac{1}{2}m_{\delta}^{\star2} \vec{\delta} ^2\Bigr].
\end{equation}

The EoS for the hyperonic neutron star matter is given by the
diagonal components of the energy-momentum tensor. Thus we have the
energy density as follows:
\begin{eqnarray}\label{eq:17}
\epsilon&=&-i\sum_{B}\int\frac{{\rm d}^{4}k}{(2\pi)^4}{\rm Tr}[\gamma^{0}G^{B}(k)]k^{0}\nonumber \\
&&+\sum_{l}\frac{1}{8\pi^2}\Bigl[k_{F_{l}}E_{F_{l}}^{\star}(m_{l}^{2}+2k_{F_{l}}^{2})
  -m_{l}^{4}{\rm ln}\Bigl(\frac{k_{F_{l}}+E_{F_{l}}^{\star}}{m_{l}}\Bigr)\Bigr]\nonumber \\
&&+\frac{1}{2}m_\sigma^{\star2}\phi^2 + U(\phi)
+\frac{1}{2}m_\omega^{\star2}\omega_0^2
+\frac{1}{2}m_\rho^{\star2}b_0^2
+\frac{1}{2}m_\delta^{\star2}\delta_{3}^{2},
\end{eqnarray}
and the pressure is given by
\begin{eqnarray}\label{eq:18}
 P &=&-\frac{i}{3}\sum_{B,i}\int\frac{{\rm d}^{4}k}{(2\pi)^4}{\rm Tr}[\gamma^{i}G^{B}(k)]k^{i}\nonumber \\
  &&+\sum_{l}\frac{1}{8\pi^2}\Bigl[m_{l}^{4}{\rm ln}\Bigl(\frac{k_{F_{l}}
  +E_{F_{l}}^{\star}}{m_{l}}\Bigr)-E_{F_{l}}^{\star}k_{F_{l}}\Bigl(m_{l}^{2}-\frac{2}{3}k_{F_{l}}^{2}\Bigr)\Bigr]\nonumber\\
&&-\frac{1}{2}m_\sigma^{\star2}\phi^2 - U(\phi)
+\frac{1}{2}m_\omega^{\star2}\omega_0^2
+\frac{1}{2}m_{\rho}^{\star2}{b_0^2}
-\frac{1}{2}m_\delta^{\star2}\delta_{3}^{2},
\end{eqnarray}
where the sum on $i$ is over the space components of $\gamma$ and
$k$, $k_{F_{l}}$ is the Fermion momentum of free leptons and
$E_{F_{l}}=\sqrt{m_{l}^2+k_{F_{l}}^2}$.

Hyperonic neutron stars are neutral charged objects in $\beta$
equilibrium. For the hyperonic neutron star matter, the chemical
potential of baryon octet states and leptons are constrained by the
baryon number and electric charge conservation:
\begin{eqnarray}\label{eq:19-23}
\mu_{\mu}&=&\mu_{e}, \\
\mu_p&=&\mu_n-\mu_{e},\\
\mu_\Lambda&=&\mu_{\Sigma^{0}}=\mu_{\Xi^{0}}=\mu_n,\\
\mu_{\Sigma^{-}}&=&\mu_{\Xi^{-}}=\mu_n+\mu_e,\\
\mu_{\Sigma^{+}}&=&\mu_p=\mu_n-\mu_e,
\end{eqnarray}
where $\mu_{n}$ and $\mu_{e}$ are the independent neutron and
electron chemical potentials, respectively, where the chemical
potentials of the baryon $B$ and the lepton $l$ are given by,
respectively,
\begin{eqnarray}\label{eq:24-25}
\mu_B&=&\sqrt{k_{B}^2+M_{B}^{\star 2}} +g_{\omega B}{\omega_0}
+g_{\rho B}{t_{3B}}{b_0},\\
{\mu_{l}}&=&\sqrt{k_{F_{l}}^2+{{m_{l}}}^2}.
\end{eqnarray}

The neutral charged condition of the hyperonic neutron star matter
can be expressed as:
\begin{equation}\label{eq:26}
\rho_{p}+\rho_{\Sigma^{+}}-\rho_{\Sigma^{-}}-\rho_{\Xi^{-}}=\rho_{e^{-}}+\rho_{\mu^{-}}.
\end{equation}

The properties of hyperonic neutron stars can be obtained by solving
Tolmann-Oppenheimer-Volkov (TOV) equations \cite{tol} with the
derived EoS as the input.

\section*{III. Results and discussions}

In this work, the meson-nucleon coupling constants are fixed by the
same saturation properties of the nuclear matter as in Ref.
\cite{guo}. In general, the interactions between different meson and
hyperon states should be different. We prefer to adopt the hyperon
potentials to determine the $\sigma$ meson-hyperon coupling
constants with the vector and isovector meson-hyperon couplings
fixed by SU(6) quark symmetry because such meson-hyperon coupling
choice can reflect the different interactions between meson and
hyperon states to some degree. For the meson-hyperon coupling
constants, it is convenient to define $x_{jH}=g_{jH}/g_{jN}$ ($N$
=$n,~p$ throughout this paper). The $\sigma$ meson-hyperon coupling
constants are fixed by the corresponding hyperon potentials,
$U_{H}=x_{\omega H}V-x_{\sigma H}S$, where $V=g_{\omega
N}\omega_{0}$ and $S=g_{\sigma N}\phi$ are the $\omega$ and $\sigma$
field strengths at the saturation density \cite{gle91,sch93}. As
discussed in Ref. \cite{mil}, $\Lambda$ is known to experience an
attractive potential, $U_{\Lambda}=-28~MeV$, in hypernuclear matter.
Recently, some authors suggested that $\Sigma^{-}$ may feel
repulsive potential at high densities \cite{fri94,fri95,fri07},
which was supported by the absence of bound states in a recent
$\Sigma$ hypernuclear search \cite{bar}. Therefore, the repulsive
potential of $\Sigma$, $U_{\Sigma}=30~MeV$, is adopted in our
calculations as in \cite{sha}. The attractive potential of $\Xi$,
$U_{\Xi}=-18~MeV$, is adopted from $\Xi$-$N$ interaction \cite{sch}.
The obtained $\sigma$ meson-hyperon coupling constants are listed in
Table I. As mentioned before, the vector and isovector meson-hyperon
couplings are fixed by SU(6) quark symmetry \cite{jur}:
\begin{eqnarray}\label{eq:27-31}
g_{\omega\Lambda}&=&g_{\omega\Sigma}=2g_{\omega\Xi}=\frac{2}{3}g_{\omega N},\\
g_{\rho\Lambda}&=&0,\\
g_{\rho\Sigma}&=&2g_{\rho\Xi}=2g_{\rho N},\\
g_{\delta\Lambda}&=&0,\\
g_{\delta\Sigma}&=&2g_{\delta\Xi}=2g_{\delta N}.
\end{eqnarray}

\par
\begin{table}
\vspace{0.3cm} \noindent {{\large \bf Table I.}~The $\sigma$
meson-hyperon coupling constants obtanied from hyperon potentials in
the VF-RMF and NL-RMF models.}
\par
\begin{center}
\vspace{0.5cm} \noindent
\begin{tabular}{c|c|c|c|c}  \hline
$parameters$    &\multicolumn{2}{|c}{$VF-RMF~model$}
               &\multicolumn{2}{|c}{$NL-RMF~model$} \\ \cline{2-5}
                        &$~~~VF\rho~~~$    &$VF\rho\delta$      &$~~NL\rho~~$     &$NL\rho\delta$ \\\hline
 $x_{\sigma\Lambda}$    &0.62              &0.64                &0.61             &0.62           \\\hline
 $x_{\sigma\Sigma}$     &0.37              &0.38                &0.36             &0.37           \\\hline
 $x_{\sigma\Xi}$        &0.33              &0.34                &0.32             &0.33           \\\hline
\end{tabular}
\end{center}
\vspace{0.5cm}
\end{table}

\begin{figure}[hbtp]\label{fig02}
\begin{center}
\includegraphics[width=10cm]{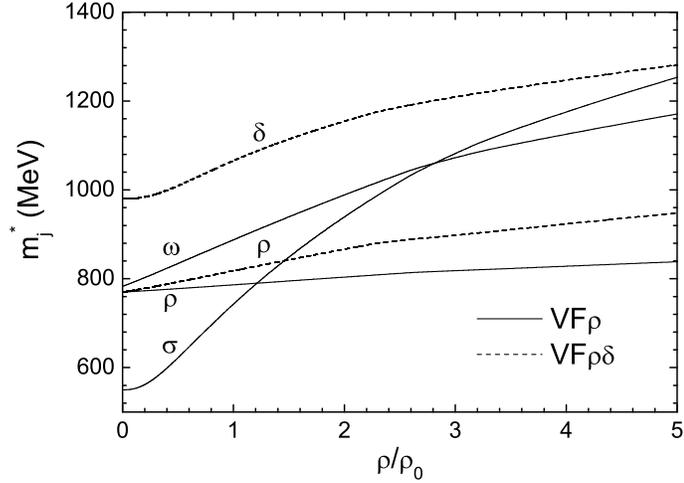}
\caption{Off-shell in-medium meson masses ($m_{j}^{\star}$) in the
hyperonic neutron star matter as a function of the total baryon
density with the meson-hyperon couplings fixed by hyperon potentials
and SU(6) quark symmetry in the VF-RMF model.}
\end{center}
\end{figure}

The in-medium masses of baryons and mesons can be obtained by
calculating the loop corrections to their self-energies, see Fig. 1.
As discussed before, because the meson propagators appearing in the
baryon self-energies are calculated at zero four-momentum transfer
[see Fig. 1(a)], we have to use the off-shell in-medium meson masses
in the tadpole loop calculations for self consistency. The off-shell
in-medium meson masses in the hyperonic neutron star matter are
shown in Fig. 2. It is found that the off-shell in-medium meson
masses increase with the increase of the total baryon density (the
sum of the densities of $n$, $p$, $\Lambda$, $\Sigma^{-}$,
$\Sigma^{0}$, $\Sigma^{+}$, $\Xi^{-}$, $\Xi^{0}$). The introduction
of the density dependence of in-medium meson masses is the critical
effect of VF corrections.

\begin{figure}[hbtp]\label{fig03}
\begin{center}
\includegraphics[width=10cm]{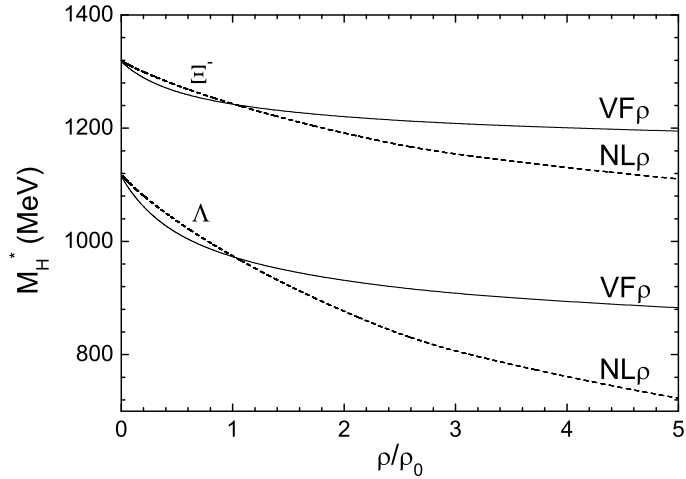}
\caption{The in-medium masses of hyperons, $M_{H}^{\star}$ ($H
=\Lambda,~\Xi^{-}$), as a function of the total baryon density with
the meson-hyperon couplings by hyperon potentials and SU(6) quark
symmetry in different models.}
\end{center}
\end{figure}

In this work, hyperons are included in the VF-RMF model. The
in-medium masses of hyperons play an important role in the
calculations of EoS. Fig. 3 shows the in-medium masses of $\Lambda$
and $\Xi^{-}$ as a function of the total baryon density in different
models for a comparison. It is found that the VF corrections soften
the decrease of the in-medium hyperon masses at high densities. This
implies the softer EoS for the hyperonic neutron star matter
obtained in the VF-RMF model than that in the NL-RMF model.

\begin{figure}[hbtp]\label{fig04}
\begin{center}
\includegraphics[width=10cm]{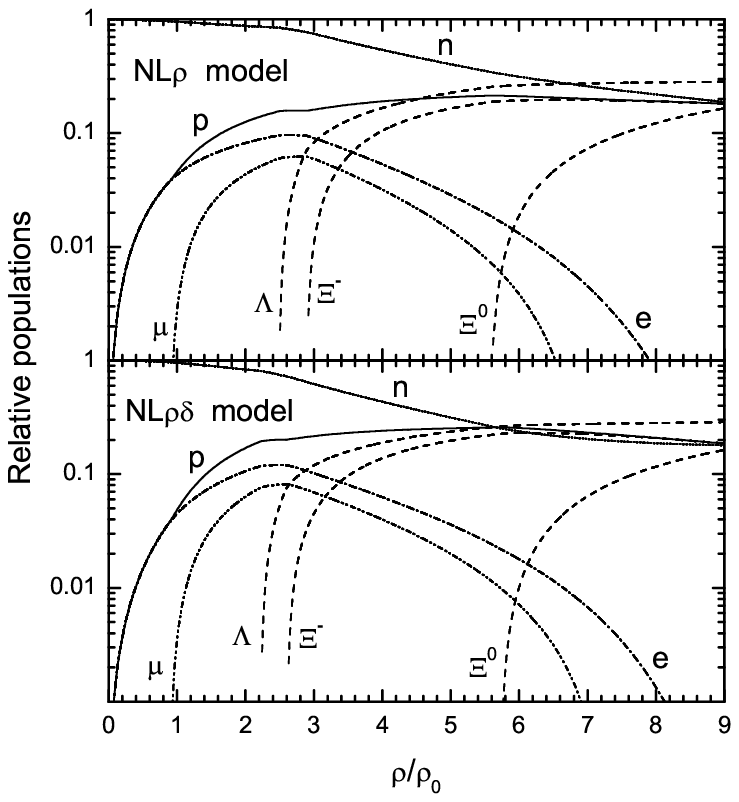}
\caption{The relative populations as a function of the total baryon
density with the meson-hyperon couplings fixed by hyperon potentials
and SU(6) quark symmetry in the NL-RMF model.}
\end{center}
\end{figure}

\begin{figure}[hbtp]\label{fig05}
\begin{center}
\includegraphics[width=10cm]{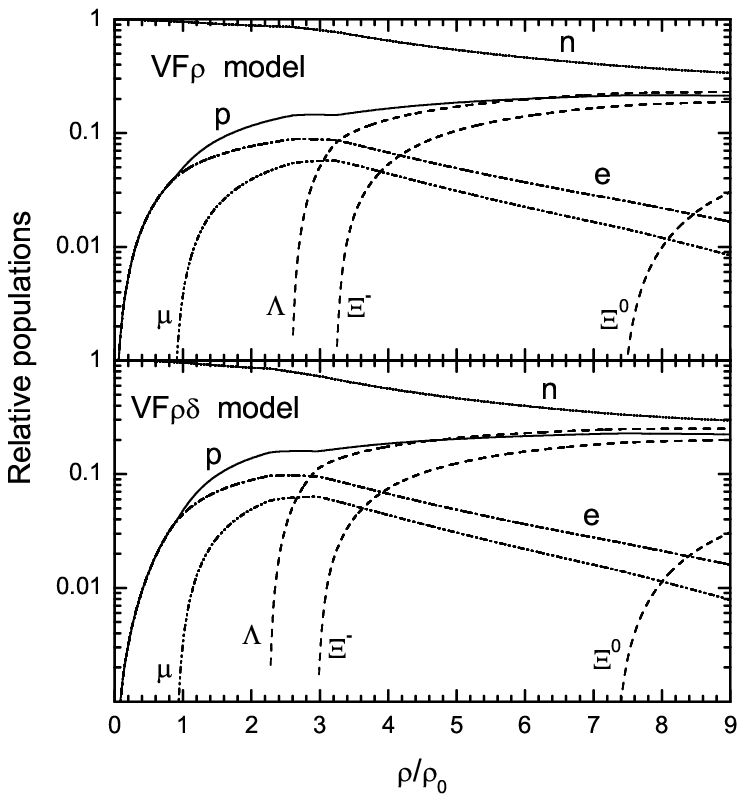}
\caption{The relative populations as a function of the total baryon
density with the meson-hyperon couplings fixed by hyperon potentials
and SU(6) quark symmetry in the VF-RMF model.}
\end{center}
\end{figure}

Now we define the relative population of the baryon $B$ as the ratio
of the density of $B$ and the total baryon density. Fig. 4 and Fig.
5 show the relative populations as a function of the total baryon
density with the meson-hyperon couplings fixed by hyperon potentials
and SU(6) quark symmetry in the NL-RMF model and the VF-RMF model,
respectively. Comparing these two figures, we find that the VF
corrections lead to later emergence of all hyperons. Meanwhile, the
VF corrections reduce the population of $\Xi^{0}$ while they enlarge
the population of leptons. We can see that $\Sigma^{(\pm,0)}$
experience such a strong repulsion that they do not appear at all in
the density range found in the neutron stars. This is consistent
with that $\Sigma^{-}$ is hardly stabilized in the hypernuclear
matter \cite{sha}. In the VF-RMF model, the $\delta$ field effects
shift the thresholds of all the hyperons to lower densities. On the
other hand, the $\delta$ field effects shift the thresholds of
$\Lambda$ and $\Xi^{-}$ to lower densities while shifting the
threshold of $\Xi^{0}$ to higher density in the NL-RMF model. The
$\delta$ field effects on the population of baryons and leptons are
not apparent in both models.

\begin{figure}[hbtp]\label{fig06}
\begin{center}
\includegraphics[width=13cm]{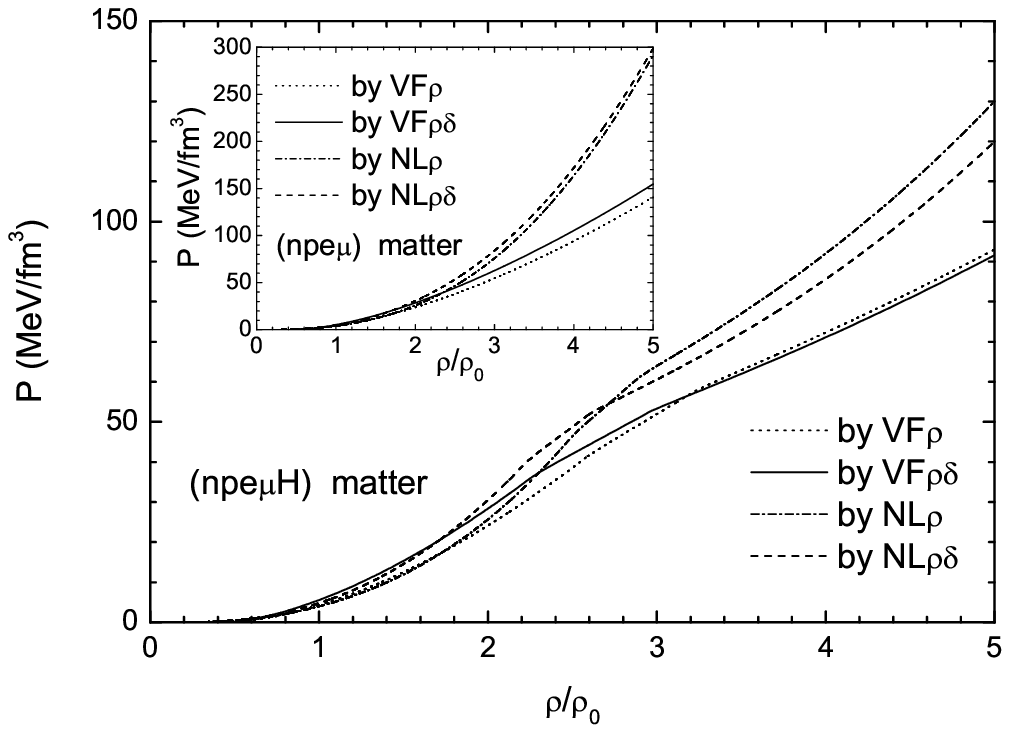}
\caption{The EoS for the hyperonic neutron star matter with the
meson-hyperon couplings fixed by hyperon potentials and SU(6) quark
symmetry. The insert is the EoS for the nucleonic neutron star
matter.}
\end{center}
\end{figure}

Fig. 6 shows the EoS, pressure vs. the total baryon density, for the
hyperonic neutron star matter with the meson-hyperon couplings fixed
by hyperon potentials and SU(6) quark symmetry in different models.
The insert of Fig. 6 presents the EoS for the nucleonic ($npe\mu$)
neutron star matter for a comparison. We can see that the
introduction of hyperons and VF corrections soften the EoS greatly.
Unlike the case of the nucleonic neutron star matter, the presence
of the $\delta$ field stiffen the EoS at first, and then from the
appearance of $\Lambda$ and $\Xi^{-}$ till higher densities soften
the EoS for the hyperonic neutron star matter. This is because that
the attractive effects of $\Lambda$ and $\Xi^{-}$ are larger than
the repulsive effects of  the $\delta$ field. Such effects of the
$\delta$ field reflect the complicated nature of interactions
between mesons and hyperons in the hyperonic neutron star matter,
which needs deeper study in the future.

\begin{figure}[hbtp]\label{fig07}
\begin{center}
\includegraphics[width=10cm]{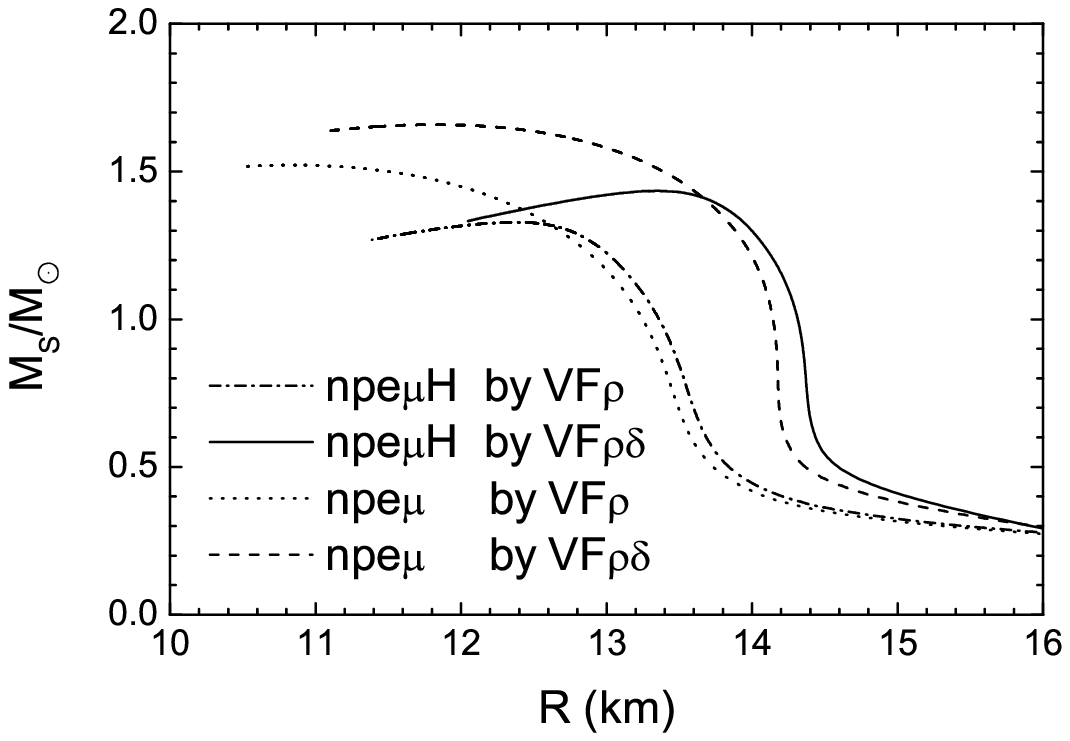}
\caption{The masses of neutron stars as a function of the radii of
neutron stars in the VF-RMF model. The meson-hyperon couplings are
fixed by hyperon potentials and SU(6) quark symmetry. }
\end{center}
\end{figure}

\begin{table}
\begin{center}
{{\large \bf Table II.}~~The maximum masses ($M_{S}$ in unit of
$M_{\bigodot}$), the corresponding radii and the central densities
of hyperonic neutron stars and nucleonic neutron stars. The
meson-hyperon couplings are fixed by hyperon potentials and SU(6)
quark symmetry.}
\par
\vspace{0.5cm} \noindent
\begin{tabular}{c|c|c|c|c|c} \hline
$        $      &     &\multicolumn{2}{|c|}{$VF-RMF~model$}
&\multicolumn{2}{c}{$NL-RMF~model$}\\
\hline $neutron~star$
&$properties$                              &$~~~VF\rho~~~$   &$VF\rho\delta$    &$~~~NL\rho~~~$      &$NL\rho\delta$ \\
\hline $hyperonic~neutron~star$
                 &$M_{S}/M_{\bigodot}$     &1.33             &1.43              &1.57                &1.59        \\ \cline {2-6}
$npe\mu H$
                 &$R (km)$                 &12.37            &13.13             &12.02               &13.036      \\  \cline {2-6}
                 &$\rho_c/\rho_0$          &4.91             &4.89              &5.76                &4.63        \\  \hline
$nucleonic~neutron~star$
                &$M_{S}/M_{\bigodot}$      &1.52             &1.66              &2.09                &2.12        \\ \cline {2-6}
$npe\mu$
                &$R (km)$                  &10.85            &11.82             &10.93               &11.37       \\ \cline {2-6}
                &$\rho_c/\rho_0$           &7.58             &6.36              &6.66                &6.29        \\ \hline
\end{tabular}
\end{center}
\end{table}

The properties of neutron stars can be calculated by solving TOV
equations. Fig. 7 shows the correlation between the neutron star
masses and the corresponding radii for hyperonic and nucleonic
neutron stars with meson-hyperon couplings obtained by hyperon
potentials and SU(6) quark symmetry in different models. The
obtained maximum masses, the corresponding radii and the central
densities are presented in Table II. As pointed out in Ref.
\cite{gle01}, the introduction of hyperons leads to the reduction of
the maximum neutron star masses due to the Pauli principle effects.
We can see from Table II that our results are consistent with this
statement. Furthermore, we can see that the VF corrections also
result in the reduction of the maximum masses of neutron stars. The
presence of the $\delta$ field enlarges the maximum masses and radii
of neutron stars.

In literatures, there are various approaches to determine the
meson-hyperon couplings \cite{gle85,jur,mos}. In order to see the
dependence of our results on the meson-hyperon couplings, we also
compare our results by adopting the meson-hyperon couplings derived
from the quark counting method, $x_{jH}=\sqrt{2/3}$ \cite{mos}, and
the universal meson-hyperon couplings, $x_{jH}=1$ \cite{gle85}.

\begin{figure}[hbtp]\label{fig08}
\begin{center}
\includegraphics[width=10cm]{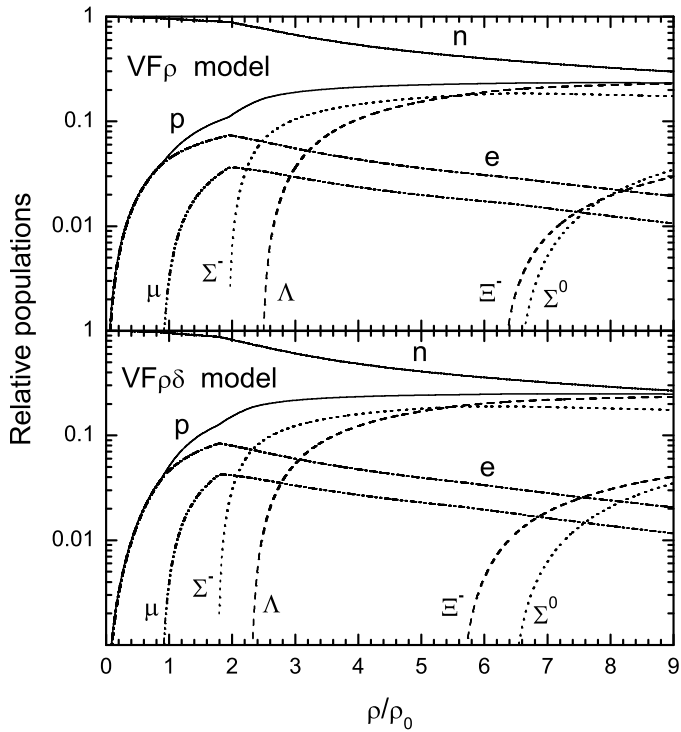}
\caption{The relative populations as a function of the total baryon
density with quark counting meson-hyperon couplings in the VF-RMF
model.}
\end{center}
\end{figure}

\begin{figure}[hbtp]\label{fig09}
\begin{center}
\includegraphics[width=10cm]{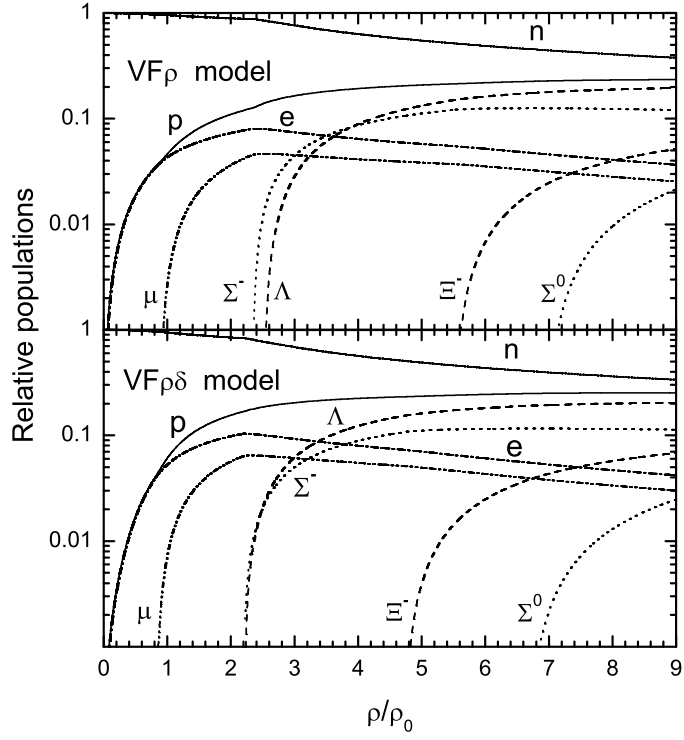}
\caption{The relative populations as a function of the total baryon
density with the universal meson-hyperon coupling choice in the
VF-RMF model.}
\end{center}
\end{figure}

 Fig. 8 and Fig. 9 show the relative populations as a function of
the total baryon density with the quark counting and universal
meson-hyperon coupling choices, respectively. We can see that
$\Sigma^{-}$ is the first hyperon to appear due to its low mass and
favored charge \cite{rio}. For both of  the two kinds of
meson-hyperon couplings, the presence of $\delta$ field leads to
earlier emergence of hyperons. For both the quark counting and the
universal meson-hyperon coupling choices, the populations of nucleons
dominate in the whole density region. However, as it can be seen from
Fig. 5, the population of $\Lambda$ exceeds that of the proton at
high densities when we adopt the meson-hyperon couplings fixed by
hyperon potentials and SU(6) quark symmetry. Comparing Fig. 8 and
Fig. 9 with Fig. 5, we can see that the onset of $\Xi^{-}$ is
greatly reduced in Fig. 5 to compensate the absence of $\Sigma^{-}$
 in order to keep charge neutrality \cite{gle85}. We note that
$\Xi^{0}$ and $\Sigma^{+}$ will appear beyond the maximum density
considered here, $9\rho_{0}$, when we adopt the quark counting and
universal meson-hyperon coupling choices.

\begin{figure}[hbtp]\label{fig10}
\begin{center}
\includegraphics[width=10cm]{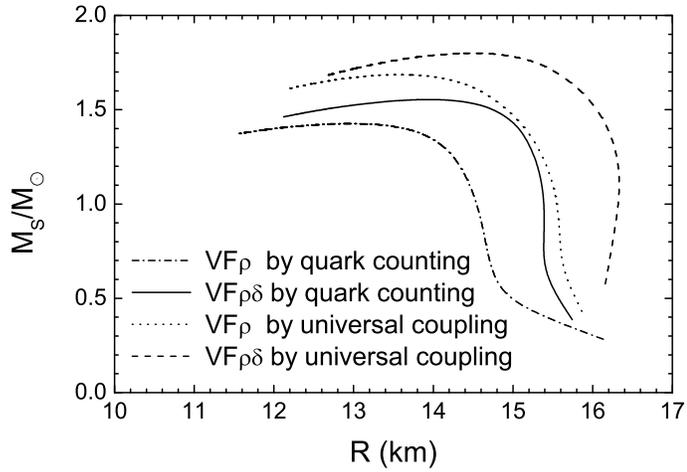}
\caption{The masses of hyperonic neutron stars as a function of the
radii of hyperonic neutron stars with the quark counting and
universal meson-hyperon couplings in the VF-RMF model. }
\end{center}
\end{figure}

Fig. 10 shows the masses of hyperonic neutron stars as a function of
the neutron star radii with the quark counting and universal
meson-hyperon couplings in the VF-RMF model. Comparing Fig. 10 with
Fig. 7, it is obvious that the properties of hyperonic neutron stars
are sensitive to the meson-hyperon couplings. We can see that the
weaker the meson-hyperon couplings the lower the masses and radii of
hyperonic neutron stars are obtained. As in Fig. 7, the presence of
the $\delta$ field also increases the masses and radii of hyperonic
neutron stars.

\begin{table*}
\begin{center}
{{\large \bf Table III.}~~The maximum mass ($M_{S}$ in unit of
$M_{\bigodot}$), the corresponding radii and the central densities
of hyperonic neutron stars with the quark counting and universal
meson-hyperon coupling choices in the VF-RMF model.}
\par
\vspace{0.5cm} \noindent

\begin{tabular}{c|c|c|c} \hline
$        $      &     &\multicolumn{2}{|c}{$VF-RMF~model$}
\\ \hline $meson-hyperon~couplings$ &$properties$  &$~~~VF\rho~~~$  &$VF\rho\delta$
\\ \hline
                     &$M_{S}/M_{\bigodot}$   &1.43            &1.55
\\\cline{2-4}
$x_{jH}=\sqrt{2/3}$  &$R (km)$               &12.98           &13.98
\\ \cline {2-4}
                     &$\rho_c/\rho_0$        &4.71            &3.98
\\ \hline
                     &$M_{S}/M_{\bigodot}$   &1.69            &1.80
\\ \cline {2-4}
$x_{jH}=1$           &$R (km)$               &13.59           &14.50
\\ \cline {2-4}
                     &$\rho_c/\rho_0 $       &4.48            &3.89
\\ \hline
\end{tabular}
\end{center}
\end{table*}

Table III displays the properties of hyperonic neutron stars with
the quark counting and universal meson-hyperon couplings
 in the VF-RMF model. Comparing Table III with Table II, we can see that
the influence of meson-hyperon couplings on our results are
distinct. With the adoption of the universal meson-hyperon
couplings, the obtained maximum hyperonic neutron star masses are
higher than those obtained by adopting the other two kinds of
meson-hyperon couplings and even higher than the maximum masses
obtained in the case of nucleonic neutron stars. As discussed in
Refs. \cite{gle91,gle01}, the conversion of nucleons to hyperons are
energetically favored. The inclusion of hyperons softens the EoS and
consequently reduces the maximum masses of neutron stars because
Pauli principle minimizes the total energy at a given density. Based
on the above discussion, the choice of universal hyperon couplings
is not appropriate for our model. It is found that the weaker the
meson-hyperon couplings the lower maximum masses and the
corresponding radii of hyperonic neutron stars are obtained. The
adoption of meson-hyperon couplings fixed by the hyperon potentials
and SU(6) quark symmetry results in the softest EoS of the hyperonic
neutron matter and hence leads to the lowest maximum masses of
hyperonic neutron stars. We also find that the maximum masses and
radii of hyperonic neutron stars increase when the $\delta$ field
presents in the VF-RMF model.

\section*{IV. Summary}

In this work, we investigate the properties of the hyperonic neutron
star matter in the extended VF-RMF model. The interactive hyperons
and free leptons are introduced into the relativistic Lagrangian
density. The VF effects are included by taking into account the loop
corrections in the hadron self-energies. In our calculations, we
replace all the meson masses by the off-shell in-medium meson masses
since the propagators of mesons in the tadpole diagrams of baryon
self-energies carry zero-momenta. With the VF corrections the
in-medium baryon and meson masses are dependent on the total baryon
density.

In general, the interactions between different mesons and hyperons
should be different. We prefer to adopt the $\sigma$ meson-hyperon
couplings derived from the hyperon potentials with the vector and
isovector meson-hyperon couplings fixed by the SU(6) quark symmetry
because this choice for the meson-hyperon couplings can reflect such
differences to a certain degree. We find that the off-shell
in-medium meson masses increase with the increase of total baryon
density. The in-medium masses of hyperons decrease slower at high
densities when the VF corrections are introduced. The density
dependence of off-shell in-medium meson masses and in-medium hyperon
masses influences the properties of the EoS for the hyepronic
neutron star matter directly. The introduction of hyperons soften
the EoS since Pauli principle minimizes the total energy at a given
density. The results obtained in the VF-RMF model are compared with
those obtained in the NL-RMF model. It is found that the VF
corrections soften the EoS for the hyperonic neutron star matter and
hence the maximum masses of hyperonic neutron stars are reduced.
$\Sigma^{(\pm,0)}$ are absent in the range of densities found in
neutron stars because they feel a strong repulsion in this
meson-hyperon couplings choice.

Then, the dependence of our results on the meson-hyperon couplings
is studied. We use other two choices for meson-hyperon couplings for
a comparison: one is derived from the quark counting method, the
other is the so-called universal couplings. It is found that the
properties of hyperonic neutron stars are sensitive to the
meson-hyperon couplings. The weaker the meson-hyperon couplings the
lower the maximum masses of hyperonic neutron stars are obtained.
For the quark counting and universal coupling choices,
$\Sigma^{(\pm,0)}$ are present and the populations of nucleons
dominant in the whole region of total baryon densities. The maximum
masses of hyperonic neutron stars obtained with the universal
meson-hyperon couplings exceed the maximum masses of nucleonic
neutron stars. Because Pauli principle assures that the appearance
of hyperons will lower the Fermi energy of baryons and hence will
lower the total energy at a given baryon density, the universal
meson-hyperon couplings is not appropriate for our model. Taking
into account the uncertainty of meson-hyperon couplings, the
obtained maximum hyperonic neutron star masses are in the range of
$1.33M_{\odot}\sim1.55M_{\odot}$ ($M_{\odot}$ denotes the mass of
the sun) in the VF-RMF model.

The effects of the $\delta$ field on hyperonic neutron stars are
investigated in the VF-RMF model. Unlike the case of the nucleonic
neutron star matter, we find that the presence of the $\delta$ field
stiffen the EoS at first and then soften the EoS from the appearance
of $\Lambda$ and $\Xi^{-}$ till higher densities for the hyperonic
neutron star matter. Such effects of the $\delta$ field on the EoS
reflect the complicated interactions in the hyperonic neutron star
matter. In addition, the presence of the $\delta$ field enlarges the
maximum masses and radii of hyperonic neutron stars. The effects of
the $\delta$ field on neutron stars are more apparent when the VF
corrections are included.

As discussed in Ref. \cite{bri}, the exchange diagram contributions
only provide small corrections to the EoS for nuclear matter in the
RMF approach at high densities. We simply extend this statement to
the case of the hyperonic neutron star matter as the first step to
study the VF effects by including hyperons in our model.
Consequently, we only consider the contributions from tadpole
diagrams to the baryon self-energies in the present work. It will be
very interesting to study the VF effects on the properties of
hyperon-hyperon interaction, kaon condensation and unconfined quarks
in the core of neutron stars in the future.

\begin{acknowledgments}
  This project is supported by the National Natural
  Science Foundation of China (Project Nos. 10675022, 10535050, 10890092, and 10875160) and
  the Special Grants from Beijing Normal University.
\end{acknowledgments}

\end{document}